\documentclass[pre,tighten]{revtex4}

\usepackage{amssymb,amsmath}

\usepackage{graphicx}

\begin{document}

\title{Thermal diffusion segregation in granular binary mixtures described by the Enskog equation}
\author{Vicente Garz\'o}
\email{vicenteg@unex.es} \homepage{http://www.unex.es/eweb/fisteor/vicente/}
\affiliation{Departamento de
F\'{\i}sica, Universidad de Extremadura, E-06071 Badajoz, Spain}

\begin{abstract}
Diffusion induced by a thermal gradient in a granular binary mixture is analyzed in the context of the (inelastic) Enskog equation. Although the Enskog equation neglects velocity correlations among particles which are about to collide, it retains spatial correlations arising from volume exclusion effects and thus it is expected to apply to moderate densities. In the steady state with gradients only along a given direction, a segregation criterion is obtained from the thermal diffusion factor $\Lambda$ measuring the amount of segregation parallel to the thermal gradient. As expected, the sign of the factor $\Lambda$ provides a criterion for the transition between the Brazil-nut effect (BNE) and the reverse Brazil-nut effect (RBNE) by varying the parameters of the mixture (masses, sizes, concentration, solid volume fraction, and coefficients of restitution). The form of the phase diagrams for the BNE/RBNE transition is illustrated in detail for several systems, with special emphasis on the significant role played by the inelasticity of collisions. In particular, an effect already found in dilute gases (segregation in a binary mixture of identical masses and sizes {\em but} different coefficients of restitution) is extended to dense systems. A comparison with recent computer simulation results shows a good qualitative agreement at the level of the thermal diffusion factor. The present analysis generalizes to arbitrary concentration previous theoretical results derived in the tracer limit case.
\end{abstract}


\date{\today}
\maketitle

\section{Introduction}
\label{sec1}

Controlling the mixing/demixing of granular media containing more than one species (polydisperse systems) is a problem faced by a wide range of industries. In some cases one would like to enhance the mixing effect while in other situations it might be a desired and useful effect to separate particles of different types. Nevertheless, in spite of its practical importance, the physical mechanisms involved in the segregation phenomenon are still not completely understood \cite{K04}. This fact has motivated the development of accurate continuum models for polydisperse solid mixtures in order to offer a reliable description of the bulk behavior of these systems.

One of the most famous examples of (size) segregation in vertically vibrated mixtures is the Brazil-nut effect (BNE), where a relatively large particle (intruder) tends to climb to the top of the sample against gravity \cite{RSPS87,KJN93,DRC93,CWHB96}. On the other hand, a series of experimental works \cite{SM98,HQL01} have also observed the reverse buoyancy effect, namely, the intruder can also sink to the bottom of the granular bed under certain
conditions (the reverse Brazil-nut effect, RBNE). Several mechanisms have been proposed to explain the transition BNE/RBNE, such as for example, void filling \cite{RSPS87}, convection \cite{KJN93,LCBRD94}, inertia \cite{SM98}, and interstitial-fluid effects \cite{MLNJ01}. Among the different competing mechanisms, thermal diffusion becomes the most relevant one when the granular system is vigorously shaken. Under those conditions, the motion of grains resembles the motion of atoms or molecules in an ordinary gas and so near-instantaneous binary collisions prevail. In this case, kinetic theory properly modified to account for the inelasticity of collisions may be quite a useful tool to provide a reliable description of the kinetics and hydrodynamics of the system, and in particular to analyze segregation in mixtures.

Thermal diffusion (or thermophoresis in its single-particle manifestation \cite{GR83}) is the transport of matter caused by the presence of a thermal gradient. Due to the motion of the components of the mixture, a steady state can be reached in which the separation effect arising from thermal diffusion is balanced by the remixing effect of ordinary diffusion. As a consequence, partial separation or segregation is observed and described by the so-called thermal diffusion factor. While this phenomenon has been widely studied in ordinary gases and liquids \cite{TDF}, much less is known on thermal diffusion in the case of granular mixtures. It must be noted that for granular systems thermal diffusion can appear in vibrated systems even in the absence of an external imposed
temperature gradient, as a consequence of inelasticity. In this case (energy supplied by vertical walls), the mean kinetic energy of the grains decays away from the source of energy giving rise to a (granular) temperature gradient.

Previous theoretical attempts to describe thermal diffusion based on kinetic theory have been reported in the past few years. In the {\em low-density} regime, Serero {\em et al.} \cite{SGNT06,SNTG09} have studied the direct influence of inelasticity alone on thermal diffusion segregation in the case of near-elastic particles \cite{SGNT06} and finite degree of dissipation \cite{SNTG09}. In particular, they
find a novel effect, namely the fact that, even when the species differ only by
their respective coefficients of restitution $\alpha_{ij}$, they may segregate when subject to a
temperature gradient. However, they assume energy equipartition, which can only be considered as acceptable when $\alpha_{ij}\simeq 1$. In fact, the failure of energy equipartition in granular mixtures \cite{GD99} has been widely confirmed by computer simulations \cite{computer} and observed in real experiments \cite{exp} of agitated mixtures. Additional efforts for dilute granular mixtures have been made to assess the impact of the breakdown of energy equipartition on thermal diffusion \cite{BRM05,G06}. Interestingly, nonequipartition plays an important role since those results show that the relative position of the large particles 1 with respect to the small particles 2 is given by the {\em sign} of the control parameter $(m_2T_1/m_1T_2)-1$, where $m_i$ and $T_i$ are the mass and partial temperature of species $i$. While in an ordinary gas this sign is fixed only by the mass ratio (since $T_1=T_2$), for a granular gas it also depends on the temperature ratio $T_1/T_2$ because of the lack of equipartition. This segregation criterion compares well with molecular dynamics simulations \cite{BRM05,SUKSS06}.

In the case of {\em dense} granular mixtures, Arnarson and Willits \cite{AW98} have determined the thermal diffusion factor for nearly elastic mixtures. However, their theory (which is based on the results of Jenkins and Mancini \cite{JM89}) differs from the theory for elastic particles \cite{KCM87} only in the fact that it includes a sink term in the equation for the temperature and so no other inelastic effects are accounted for. Slightly different approaches \cite{HH96} based on kinetic theory have been invoked to get a segregation criterion \cite{JY02,TAH03,YJ06} in the absence of a temperature gradient. In this latter case, the segregation dynamics of the intruder is only driven by the gravitational force.

The purpose of this paper is to determine the thermal diffusion factor $\Lambda$ of a {\em moderately} dense granular binary mixture described by the (inelastic) Enskog equation. Since the main interest here lies in the analysis of the effect of a thermal gradient on granular segregation, it will be assumed that no body forces (e.g., gravity) are present in the system. The segregation criterion is obtained from the factor $\Lambda$, which is explicitly given in terms of the parameters of the system (masses and sizes of particles, concentration, solid volume fraction, and coefficients of restitution). More specifically, the sign of $\Lambda$ determines the tendency of the large particles to drift toward the cooler or warmer plate. It is apparent that the knowledge of the thermal diffusion factor allows one to analyze the origin of its sign and how it is related to the different parameters of the system. Previous theoretical results \cite{G08,GV09,G09} on thermal diffusion have been recently reported by the author of the present paper in the intruder limit case when the gas is driven by an external thermostat. The objective here is to extend the above results to arbitrary concentration and compare these theoretical results with some recent molecular dynamics simulations \cite{GDH05} of a granular segregating binary system subjected to a temperature gradient.

It must remarked that the factor $\Lambda$ has been obtained from a solution \cite{GDH07,GHD07} of the Enskog equation that goes beyond the quasielastic limit (and thus, it applies for a wide range of values of the coefficients of restitution) and takes into account the non-equipartition of kinetic energy. In this context, our theory subsumes all previous studies for both dilute \cite{BRM05,G06} and dense \cite{AW98,HH96,JY02,TAH03} mixtures and, additionally assesses the influence of concentration on thermal diffusion without any restriction on the parameter space (for comprehensive review for mixture theories, see \cite{H011}). This is the main added value in this paper since our results can be relevant for comparison with experiments/simulations at finite densities. Moreover, it must be stated that the Navier-Stokes hydrodynamic equations are not actually solved in the present paper. Instead the results are conditional: if the temperature gradient has a given form, then the segregation criterion has a resultant form. For instance, in the case of a temperature inversion the results for segregation reverse.

The plan of the paper is as follows. In Section \ref{sec2} the thermal diffusion factor $\Lambda$ is defined and evaluated by using a hydrodynamic description. The factor $\Lambda$ is expressed in terms of the pressure $p$ and the transport coefficients $D_1^{T}$, $D_{11}$, and $D_{12}$ associated with the mass flux. All these coefficients have been explicitly determined from a Chapman-Enskog solution \cite{GDH07,GHD07} of the Enskog kinetic equation. The explicit forms of the pressure and the transport coefficients are displayed in Appendix \ref{appA}. The knowledge of $p$, $D_1^{T}$, $D_{11}$, and $D_{12}$ allows one to determine the thermal diffusion $\Lambda$ as a function of the parameter space of the system: the mass ($m_1/m_2$) and diameter ($\sigma_1/\sigma_2$) ratios, the concentration $x_1$, the solid volume fraction $\phi$, and the three independent coefficients of restitution of the binary mixture $\alpha_{ij}$. In order to assess the impact of the different parameters on the segregation criterion, some special situations are analyzed and illustrated with detail in Section \ref{sec3}. In Section \ref{sec4}, the form of the phase diagrams BNE/RBNE in the   $\{\sigma_1/\sigma_2, m_1/m_2\}$-plane is widely investigated by varying the parameters of the system in the case of a common coefficient of restitution ($\alpha_{11}=\alpha_{22}=\alpha_{12}\equiv \alpha$). Section \ref{sec5} deals with the comparison between the Enskog theory and molecular dynamics results \cite{GDH05} for the thermal diffusion factor. The paper is closed in Section \ref{sec6} with a brief discussion of the results.

\section{Enskog kinetic theory for thermal diffusion}
\label{sec2}

We consider a binary mixture of inelastic hard disks ($d=2$) or spheres ($d=3$) of
masses $m_i$ and diameters $\sigma_i$ ($i=1,2$). Without loss of generality, we assume
that $\sigma_1>\sigma_2$. The inelasticity of collisions among all pairs is
characterized by three independent constant coefficients of restitution $\alpha _{11}$,
$\alpha _{22}$, and $\alpha _{12}=\alpha _{21}$. For moderate densities, it is assumed that the velocity distribution functions $f_i({\bf r}, {\bf v}; t)$ of each species are accurately described by the coupled set of {\em inelastic} Enskog equations \cite{GS95,BDS97}. Like the Boltzmann equation, the Enskog equation neglects velocity correlations among particles which are about to collide, but it takes into account the dominant spatial correlations due to excluded-volume effects.
In the hydrodynamic description, it is assumed that the state of the mixture is characterized by the local number densities $n_i({\bf r},t)$, the flow velocity ${\bf U}({\bf r},t)$, and the (total) {\em granular} temperature $T({\bf r},t)$. These hydrodynamic fields are defined in terms of the velocity distribution functions $f_i$ as
\begin{equation}
\label{2.n1}
n_i=\int\; d{\bf v}  f_i({\bf v}), \quad
\rho{\bf U}=\sum_{i}m_i\int\; d{\bf v}  {\bf v}f_i({\bf v}),
\end{equation}
\begin{equation}
\label{2.n2}
nT=\sum_{i}\frac{m_i}{d}\int\; d{\bf v}  V^2 f_i({\bf v}),
\end{equation}
where $\rho=\sum_i m_in_i$ is the total mass density, $n=\sum_i n_i$ is the total number density, and ${\bf V}={\bf v}-{\bf U}$ is the peculiar velocity. Assuming that there are no external forces acting on the mixture, the macroscopic balance equations for mass, momentum and energy can be derived from the Enskog equation. They are given by \cite{GDH07}
\begin{equation}
D_{t}n_{i}+n_{i}\nabla \cdot {\bf U}+\frac{\nabla \cdot {\bf j}_{i}}{m_{i}} =0\;,
\label{2.1}
\end{equation}
\begin{equation}
D_{t}{\bf U}+\rho ^{-1}\nabla \cdot {\sf P}={\bf 0}\;, \label{2.2}
\end{equation}
\begin{equation}
D_{t}T-\frac{T}{n}\sum_{i}\frac{\nabla \cdot {\bf j}_{i}}{m_{i}}+\frac{2}{dn}
\left( \nabla \cdot {\bf q}+{\sf P}:\nabla {\bf U}\right) =-\zeta \,T. \label{2.3}
\end{equation}
Here, $D_{t}=\partial _{t}+{\bf U}\cdot \nabla $ is the material derivative, ${\bf j}_i$ is the mass
flux of species $i$, ${\sf P}$ is the pressure tensor, ${\bf q}$ is the heat flux, and
$\zeta$ is the cooling rate associated with the energy dissipation in collisions.

The constitutive equations for the irreversible fluxes ${\bf j}_i$, ${\sf P}$, and ${\bf q}$, and the cooling rate $\zeta$ have been recently obtained up to the Navier-Stokes (NS) order (first order in the spatial gradients) from the Chapman-Enskog solution \cite{CC70} to the Enskog equation \cite{GDH07,GHD07}. The results are
\begin{equation}
\mathbf{j}_{1}=-\frac{m_1^2n_{1}}{\rho} D_{11}\nabla \ln n_{1}-
\frac{m_1 m_2n_{2}}{\rho} D_{12}\nabla \ln n_{2}
-\rho D_{1}^{T}\nabla \ln T, \quad {\bf j}_2=-{\bf j}_1, \label{2.4}
\end{equation}
\begin{equation}
\label{2.5}
P_{\alpha\beta}=p\delta_{\alpha\beta}-\eta\left(\nabla_\alpha U_\beta+
\nabla_\beta U_\alpha
-\frac{2}{d} \nabla\cdot{\bf U}\;\delta_{\alpha\beta}\right)
-\kappa \nabla\cdot{\bf U}\;\delta_{\alpha\beta}
\end{equation}
\begin{equation}
\label{2.6}
{\bf q}=-T^2D_{q,1}\nabla \ln n_1-T^2 D_{q,2}\nabla \ln n_2-\lambda
\nabla T,
\end{equation}
\begin{equation}
\label{2.7}
\zeta=\zeta^{(0)}+\zeta_u\bf{\nabla}\cdot{\bf U}.
\end{equation}
In these equations, $D_{ij}$ are the mutual diffusion coefficients, $D_1^T$ is the thermal diffusion coefficient, $p$ is the pressure, $\eta$ is the shear viscosity, $\kappa$ is the bulk viscosity, $D_{q,i}$ are the Dufour coefficients, $\lambda$ is the thermal conductivity coefficient, $\zeta^{(0)}$ is the zeroth-order cooling rate and $\zeta_u$ is a transport coefficient associated with first-order cooling rate. All the above quantities have been explicitly obtained by considering the leading terms in a Sonine polynomial expansion \cite{GHD07}.

\subsection{Thermal diffusion factor}

As said in the Introduction, we are interested in analyzing segregation by thermal diffusion in a binary mixture. The amount of segregation parallel to the thermal gradient may be characterized by the thermal diffusion factor $\Lambda$. This quantity is defined in an inhomogeneous non-convecting (${\bf U}={\bf 0}$) steady state with zero mass flux (${\bf j}_1={\bf 0}$) through the relation
\begin{equation}
\label{2.8} -\Lambda\frac{\partial \ln T}{\partial z} =\frac{\partial}{\partial z}\ln
\left(\frac{n_1}{n_2}\right),
\end{equation}
where gradients only along the $z$ axis (vertical direction) have been assumed for
simplicity. Let us assume that the gas is enclosed between two plates where the bottom plate is hotter than the top plate, i.e., $\partial_z \ln T<0$.  In this geometry, according to Eq.\ (\ref{2.8}), when $\Lambda >0$ the  larger particles $1$ tend to rise with respect to the
smaller particles $2$ (i.e., $\partial_z\ln (n_1/n_2)>0$). On the other hand, when $\Lambda <0$, the
larger particles fall with respect to the smaller particles (i.e., $\partial_z\ln
(n_1/n_2)<0$). The former situation will be referred here to as the Brazil-nut effect (BNE) while
the latter will be called the reverse Brazil-nut effect (RBNE).

We obtain now a relation of the type (\ref{2.8}) from the balance equations. First, according to Eq.\ (\ref{2.4}), the steady-state condition $j_{1,z}=0$ yields
\begin{equation}
\label{2.9} -(x_1\lambda_1D_{11}^*+x_2\lambda_2D_{12}^*)=D_1^{T*},
\end{equation}
where $x_i=n_i/n$ is the mole fraction or concentration of species $i$,
\begin{equation}
\label{2.10} \lambda_i=\frac{\partial_z \ln n_i}{\partial_z \ln T} ,
\end{equation}
and we have introduced the reduced coefficients
\begin{equation}
\label{2.11} D_{ij}^*=\frac{m_im_j\nu_0}{\rho T}D_{ij},\quad
D_{1}^{T*}=\frac{\rho\nu_0}{n T}D_{1}^{T},
\end{equation}
where $\nu_0$ is an effective collision frequency defined below Eq.\ \eqref{a5}.
Moreover, since ${\bf U}={\bf 0}$, Eq.\ (\ref{2.5}) clearly shows that the pressure tensor is diagonal for this state and so, $P_{\alpha\beta}=p\delta_{\alpha\beta}$. In this case, the momentum balance equation (\ref{2.2})
reduces simply to
\begin{equation}
\label{2.12} \frac{\partial p}{\partial z}=0.
\end{equation}
The spatial dependence of the pressure $p$ is through its
dependence on the number densities $n_i$ and the temperature $T$. As a consequence,
in reduced units, Eq.\ ({\ref{2.12}) can be written more explicitly as
\begin{equation}
\label{2.13} -(x_1\beta_1\lambda_1+x_2\beta_2\lambda_2)=p^*,
\end{equation}
where $p^*=p/nT$ and
\begin{equation}
\label{2.14.1}
\beta_1=T^{-1}\frac{\partial p}{\partial n_1}=p^*+\frac{\phi_1}{x_1}\frac{\partial p^*}{\partial \phi}+x_2\frac{\partial p^*}{\partial x_1},
\end{equation}
\begin{equation}
\label{2.14.2}
\beta_2=T^{-1}\frac{\partial p}{\partial n_2}=p^*+\frac{\phi_2}{x_2}\frac{\partial p^*}{\partial \phi}-x_1\frac{\partial p^*}{\partial x_1}.
\end{equation}
Here, $\phi=\phi_1+\phi_2$ is the total solid volume fraction where $\phi_i$ is the partial solid volume fraction of species $i$ given by
\begin{equation}
\label{2.14.3}
\phi_i=\frac{\pi^{d/2}}{2^{d-1}d\Gamma (d/2)}n_i\sigma_i^d,
\end{equation}
where $\Gamma$ refers to Gamma function.
\begin{figure}
\includegraphics[width=0.5 \columnwidth,angle=0]{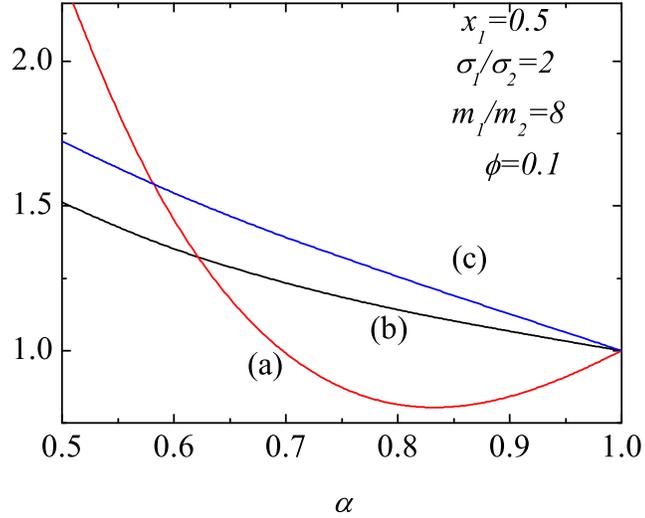}
\caption{(color online) Plot of the reduced coefficients (a) $D_{11}(\alpha)/D_{11}(1)$, (b) $D_{12}(\alpha)/D_{12}(1)$ and (c) $D_{1}^T(\alpha)/D_{1}^T(1)$ as functions of the (common) coefficient of restitution $\alpha$ for hard spheres ($d=3$) with $x_1=1/2$, $\sigma_1/\sigma_2=2$, $m_1/m_2=8$ and a solid volume fraction $\phi=0.1$.
\label{fig1}}
\end{figure}
\begin{figure}
\includegraphics[width=0.5 \columnwidth,angle=0]{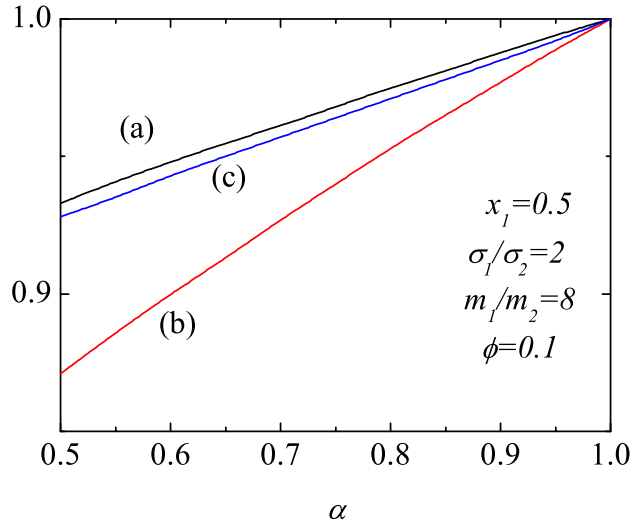}
\caption{(color online) Plot of the reduced coefficients (a) $p(\alpha)/p(1)$, (b) $\beta_{1}(\alpha)/\beta_{1}(1)$ and (c) $\beta_{2}(\alpha)/\beta_{2}(1)$ as functions of the (common) coefficient of restitution $\alpha$ for hard spheres ($d=3$) with $x_1=1/2$, $\sigma_1/\sigma_2=2$, $m_1/m_2=8$ and a solid volume fraction $\phi=0.1$.
\label{fig2}}
\end{figure}

The solution to the set of linear equations (\ref{2.9}) and (\ref{2.13}) is
\begin{equation}
\label{2.15}
\lambda_1=\frac{p^*D_{12}^*-\beta_2D_1^{T*}}{x_1(\beta_2D_{11}^*-\beta_1D_{12}^*)},
\quad
\lambda_2=\frac{\beta_1D_1^{T*}-p^*D_{11}^*}{x_2(\beta_2D_{11}^*-\beta_1D_{12}^*)}.
\end{equation}
According to Eq.\ \eqref{2.8}, $\Lambda=\lambda_2-\lambda_1$. Therefore, the thermal diffusion factor $\Lambda$ is
\begin{equation}
\label{2.16}
\Lambda=\frac{D_1^{T*}(x_1\beta_1+x_2\beta_2)-p^*(x_1D_{11}^*+x_2D_{12}^*)}{ x_1x_2(\beta_2D_{11}^*-\beta_1D_{12}^*)}.
\end{equation}
It is quite apparent that the influence of the parameters of the mixture on the sign of $\Lambda$ is rather complicated, given the large number of parameters involved. For the sake of concreteness, we consider the region of the parameter space where $\beta_2D_{11}^*-\beta_1D_{12}^*\neq 0$. In this case, the condition $\Lambda=0$ (which provides the criterion for the BNE/RBNE transition) implies [see the denominator of Eq.\ \eqref{2.16}]
\begin{equation}
\label{2.17} p^*(x_1D_{11}^*+x_2D_{12}^*)=(x_1\beta_1+x_2\beta_2)D_1^{T*}.
\end{equation}
When the parameter set yields $\beta_2D_{11}^*-\beta_1D_{12}^*=0$, $|\Lambda|\to \infty$ and BNE (RBNE) appears if $\Lambda>0$ ($\Lambda<0$). In any case, the condition $\beta_2D_{11}^*-\beta_1D_{12}^*\neq 0$ covers practically all the parameter space of the problem.

Equation \eqref{2.17} gives the curves delineating the regimes between BNE and RBNE.
To get the dependence of the thermal diffusion factor on the parameters of the mixture, the
explicit form of the transport coefficients and the equation of state is needed. These forms were evaluated in Refs.\ \cite{GDH07,GHD07} and some corrections to the expressions presented in these references have been done more recently in Ref.\ \cite{AHG10}. The final correct expressions are displayed in Appendix \ref{appA} for the sake of completeness.

Figures \ref{fig1} and \ref{fig2} show the dependence of the diffusion transport coefficients ($D_{11}$, $D_{12}$, $D_1^T$), the pressure $p$, and its derivatives $\beta_i$ with respect to the partial densities as functions of the (common) coefficient of restitution $\alpha\equiv \alpha_{ij}$. We have considered here an equimolar mixture ($x_1=1/2$) with a solid volume fraction $\phi=0.1$ and mechanical parameters  $\sigma_1/\sigma_2=2$ and $m_1/m_2=8$. To show more clearly the influence of inelasticity in collisions on mass transport and equation of state, all the quantities have been normalized with respect to their values in the elastic limit. We observe that the effect of collisional dissipation is in general significant, especially in the case of the diffusion coefficients.

\section{Some special limit situations}
\label{sec3}

The explicit form of the thermal diffusion factor $\Lambda$ can be obtained when one substitutes Eqs.\ (\ref{a1})--(\ref{a3}) for $D_{1}^{T*}$, $D_{11}^*$, and $D_{12}^*$, respectively, and Eq.\ (\ref{a10}) for $p^*$ (and its corresponding derivatives $\beta_i$) into Eq.\ (\ref{2.16}). This gives the dependence of $\Lambda$ on the parameter space of the problem (mass and size ratios, mole fraction, solid volume fraction and coefficients of restitution). It is apparent that this dependence is in general quite complex. Thus, in order to show more clearly the different competing mechanisms appearing in the segregation phenomenon, it is first convenient to consider some special situations where a more simplified criterion can be obtained.

\subsection{Mechanically equivalent particles}

This is quite a trivial case since the system is in fact monodisperse ($m_1=m_2$, $\sigma_1=\sigma_2$, $\alpha_{11}=\alpha_{22}=\alpha_{12}$). In this limit case, Eq.\ \eqref{a1}
shows that the thermal diffusion coefficient vanishes ($D_{1}^{T*}=0$), while $D_{11}^*$ and $D_{12}^*$ are given by Eqs.\ \eqref{a24}. Consequently, since the combination $x_1D_{11}^*+x_2D_{12}^*=0$, the factor $\Lambda$ vanishes [see Eq.\ (\ref{2.16})] and the condition (\ref{2.17}) holds for any value of the coefficient of restitution and volume fraction. In this case, as expected, no segregation is possible.

\subsection{Dilute binary mixtures}

Let us consider a binary mixture in the low-density regime ($\phi\to 0$ or, equivalently, $n_i\sigma_i^d\to 0$). In this regime, $p^*=\beta_i=1$, and
\begin{equation}
\label{3.0}
n_1\frac{\partial \zeta^{(0)}}{\partial n_1}+n_2\frac{\partial \zeta^{(0)}}{\partial n_2}=\zeta^{(0)},\quad
n_1\frac{\partial p}{\partial n_1}+n_2\frac{\partial p}{\partial n_2}=p.
\end{equation}
Taking into account these identities, it is easy to get the explicit form of the transport coefficients from Eqs.\ \eqref{a1}--\eqref{a3}. They can be written as
\begin{equation}
\label{3.2}
D_1^{T*}=\left(\nu_D^*-\zeta^*\right)^{-1}\left(x_1\gamma_1-
\frac{\rho_1}{\rho}\right).
\end{equation}
\begin{equation}
\label{3.1}
x_1D_{11}^*+x_2D_{12}^*=\left(\nu_D^*-\frac{1}{2}\zeta^*\right)^{-1}\left(\zeta^*D_1^{T*}+x_1\gamma_1-
\frac{\rho_1}{\rho}\right),
\end{equation}
where $\gamma_1\equiv T_1/T$ (the partial temperatures $T_i$ are in general different from the total temperature $T$) and $\nu_D^*$ and $\zeta^*$ are defined by Eqs.\ \eqref{a4} and \eqref{a5}, respectively, with $\chi_{ij}=1$. According to the above expressions, the criterion (\ref{2.17}) becomes simply
\begin{equation}
\label{3.3}
\frac{x_1x_2M\zeta^*}{(\nu_D^*-\zeta^*)(x_2+x_1\gamma)(x_2+x_1 M)}\left(
\frac{\gamma}{M}-1\right)=0,
\end{equation}
where $M\equiv m_1/m_2$ is the mass ratio and $\gamma\equiv T_1/T_2$ is the temperature ratio. Since in general $\nu_D^*>\zeta^*$, the solution to Eq.\ (\ref{3.3}) is simply
\begin{equation}
\label{3.3.0}
\frac{m_1}{m_2}=\frac{T_1}{T_2}.
\end{equation}
Although the explicit form of $\Lambda$ derived here for a (unforced) dilute mixture differs from the one obtained when the mixture is driven (heated) by means of a stochastic external force \cite{G06}, the segregation criterion \eqref{3.3.0} (based on the sign of $\Lambda$) is the same as the one found in Ref.\ \cite{G06}.   Note that if one assumes energy equipartition ($T_1=T_2$), then segregation is only predicted for particles that differ in mass, no matter what their diameters may be. It must be remarked that the condition \eqref{3.3.0} is rather complicated since it involves all the parameter space of the system.  As said in the Introduction, the criterion \eqref{3.3.0} compares well with molecular dynamics simulations \cite{BRM05} carried out in the tracer limit case ($x_1\to 0$) and is also able to explain \cite{G06} some of the molecular dynamics segregation results \cite{SUKSS06} observed in agitated mixtures constituted by particles of the same mass density and equal volumes of large and small particles.
\begin{figure}
\includegraphics[width=0.5 \columnwidth,angle=0]{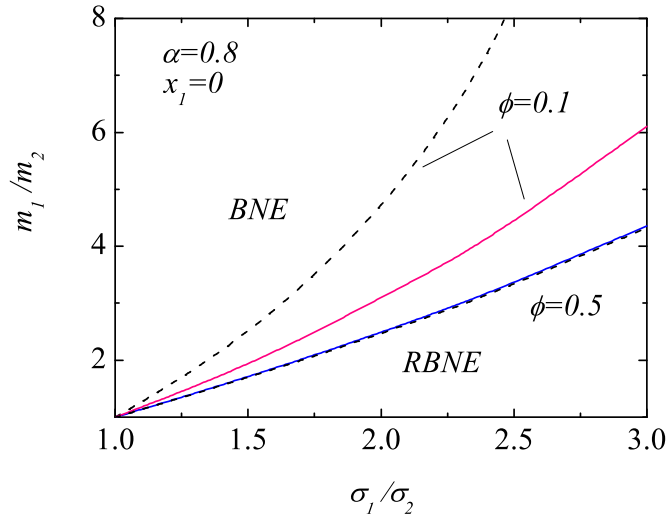}
\caption{(color online) BNE/RBNE phase diagram for inelastic hard spheres ($d=3$) with $\alpha\equiv \alpha_{12}=\alpha_{22}=0.8$ in the tracer limit case ($x_1=0$) for two different values of the solid volume fraction $\phi$. The solid lines correspond to the values derived from the relation \eqref{3.10}, while the dashed line refers to the results obtained when the gas is driven by an external thermostat, Eq.\ \eqref{b11}. Points above the curve correspond to $\Lambda>0$ (BNE) while points below the curve correspond to $\Lambda<0$ (RBNE).
\label{fig3}}
\end{figure}

\subsection{Tracer limit in a dense binary mixture}

Let us consider now a dense binary mixture where one of the components is present in tracer concentration ($x_1\to 0$). This problem is formally equivalent to studying the dynamics of an intruder immersed in a  granular gas. The tracer limit case simplifies significantly the evaluation of the transport
coefficients since, for instance, the dependence of the temperature ratio $\gamma=T_1/T_2$ on the partial densities is only through the volume fraction $\phi$ [see Eq.\ \eqref{b3}]. The explicit expressions for the diffusion coefficients in the tracer limit are given by Eqs.\ \eqref{b4}--\eqref{b6}. According to these expressions, $D_1^{T*}$ and $D_{11}^*$ are proportional to the concentration $x_1$ and so Eq.\ \eqref{2.16} for the thermal diffusion factor $\Lambda$ becomes
\begin{equation}
\label{3.3.1}
\Lambda=\frac{\beta x_1^{-1}D_1^{T*}-p^*(D_{11}^*+x_1^{-1}D_{12}^*)}{\beta D_{11}^*},
\end{equation}
where  $\beta=p^*+\phi \partial_\phi p^*$,
\begin{equation}
\label{3.7}
p^*=1+2^{d-2}\chi_{22}\phi (1+\alpha_{22})
\end{equation}
is the (reduced) pressure of the excess component and
\begin{equation}
\phi\equiv \frac{\pi^{d/2}}{2^{d-1}d\Gamma(d/2)} n_2\sigma_{2}^d \label{3.8}
\end{equation}
is the total solid volume fraction. Since $\beta$ and $D_{11}^*$ are positive in the tracer limit, then the condition $\Lambda=0$ leads to the segregation criterion
\begin{equation}
\label{3.9}
\beta D_1^{T*}=p^*(x_1D_{11}^*+x_2D_{12}^*).
\end{equation}
This criterion can be written more explicitly when one takes into account Eqs.\ \eqref{b4}--\eqref{b6} with the result
\begin{eqnarray}
\label{3.10}
& & \left[\left(\nu_D^*-\frac{1}{2}\zeta^*\right)\beta-p^*\zeta^*\left(1+\phi \partial_\phi \ln \chi_{22} \right)\right]\left[\gamma-M p^*+
\frac{(1+\omega)^d}{2}\frac{M}{1+M} \chi_{12}\phi(1+\alpha_{12})\right]
\nonumber\\
&=&p^*(\nu_D^*-\zeta^*)
\left[\gamma-M \beta +\phi \frac{\partial \gamma}{\partial \phi}+\frac{1}{2}
\frac{\gamma+M}{1+M}\frac{\phi}{T}
\left(\frac{\partial\mu_1}{\partial
\phi}\right)_{T,n_2}(1+\alpha_{12})\right],
\end{eqnarray}
where $\omega=\sigma_1/\sigma_2$ is the size ratio and $\mu_1$ is the chemical potential of the tracer particles [given by Eq.\ \eqref{a17} for $d=2$ and Eq.\ \eqref{a19} for $d=3$]. It is apparent that in spite of the tracer limit case considered, the segregation criterion (\ref{3.10}) is quite intricate and so it is not easy to disentangle the impact of each effect (nonequipartition, dissipation, density, and/or mass and size ratios) on thermal diffusion.

Thermal diffusion segregation of an intruder in a granular dense gas has been recently studied \cite{G08,G09}. In order to maintain the granular medium in a fluidized state, particles of the gas were assumed to be heated by a stochastic-driving force which mimics a thermal bath. This kind of forcing, which has been shown to be relevant for some two-dimensional experimental configurations with a rough vibrating piston \cite{PEU02}, has been used in the past by many authors \cite{thermostat} to analyze different problems, including segregation in granular mixtures \cite{TAH03}. Although previous experiments in vibrated granular mixtures \cite{exp} have shown a less significant dependence of the temperature ratio $T_1/T_2$ on inelasticity than the one obtained \cite{DHGD02} in systems heated by an external thermostat, some results (see for instance, Fig.\ 2 of Ref.\ \cite{G09}) derived for $T_1/T_2$ from this stochastic driving method compare well with molecular dynamics simulations of shaken mixtures \cite{SUKSS06}. This agreement suggests that this stochastic thermostat can be seen as a plausible approximation to modelize the experiments carried out in driven systems. On the other hand, more comparisons between results derived for driven and heated systems are needed before quantitative conclusions can be drawn on the reliability of the segregation conditions obtained from the transport coefficients derived with \cite{G09} and/or without \cite{GDH07,GHD07} an external thermostat.

As expected, the external thermostat does not play a neutral role in the transport properties of the system \cite{GM02} and, consequently, the criterion \eqref{3.10} differs from the one obtained in the driven heated case, Eq.\ \eqref{b11}. To illustrate more clearly these differences, a phase diagram delineating the regimes between BNE ($\Lambda>0$ when $\partial_zT<0$) and RBNE ($\Lambda<0$ when $\partial_zT<0$) in the $\{\sigma_1/\sigma_2, m_1/m_2 \}$-plane  is shown in Fig.\ \ref{fig3} for $\alpha\equiv \alpha_{22}=\alpha_{12}=0.8$ and two different values of the solid volume fraction $\phi$. Significant quantitative discrepancies between the predictions obtained with and without a thermostat appear for small densities ($\phi=0.1$), since the effect of the thermostat being to reduce the size of the BNE region. Much less influence appears as the density of the gas increases since both results are practically indistinguishable at higher densities ($\phi=0.5$). With respect to the influence of the solid volume fraction, we observe that in general the role played by the density of the gas is quite important since the range of size and mass ratios for which the RBNE exists increases with decreasing $\phi$.

\subsection{Inelasticity-driven segregation}
\begin{figure}
\includegraphics[width=0.5 \columnwidth,angle=0]{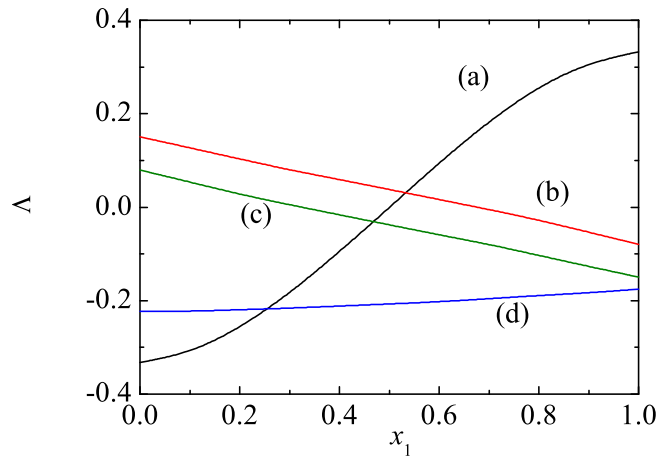}
\caption{(color online) Plot of the thermal diffusion factor $\Lambda$ as a function of the concentration $x_1$ for a solid volume fraction $\phi=0.2$ and different values of the coefficients of restitution: (a) $\alpha_{11}=\alpha_{22}=0.5$, $\alpha_{12}=0.9$; (b) $\alpha_{11}=0.8$, $\alpha_{22}=0.9$, $\alpha_{12}=0.7$; (c) $\alpha_{11}=0.9$, $\alpha_{22}=0.8$, $\alpha_{12}=0.7$; and (d) $\alpha_{11}=1$, $\alpha_{22}=0.5$, $\alpha_{12}=0.75$. Here, $m_1=m_2$ and $\sigma_1=\sigma_2$.
\label{fig4}}
\end{figure}

In a previous theoretical approach \cite{SGNT06} based on a solution of the Boltzmann equation for nearly elastic particles, it has been found that segregation is induced by inelasticity. In other words, there is a separation between both species when they differ {\em only} by their respective coefficients of restitution. The authors explain the phenomenon as a consequence of the temperature gradient induced in the system by inelastic collisions, and relate the concentration gradient with the temperature gradient. These results have been subsequently extended to arbitrary degree of inelasticity \cite{SNTG09}. The above novel effect has been also confirmed more recently \cite{BEGS08,BS09} by molecular dynamics simulations of a two-dimensional binary mixture kept fluidized by a vibrating base.

In order to study the (pure) effect of inelasticity on thermal diffusion segregation, we consider the case $m_1=m_2$ and $\sigma_1=\sigma_2$ but different coefficients of restitution $\alpha_{ij}$. Clearly, when all the coefficients of restitution are equal ($\alpha_{11}=\alpha_{22}=\alpha_{12}$), the system is monodisperse and so there is no segregation ($\Lambda=0$). Figure \ref{fig4} presents plots of the thermal diffusion factor $\Lambda$ as a function of the concentration $x_1$ for different values of the coefficients of restitution at a density $\phi=0.2$. As in the case of dilute gases ($\phi=0$) \cite{SGNT06}, segregation in the presence of a temperature gradient can then occur due to inelasticity alone. In particular, segregation occurs even if one type of collisions is elastic (case (d)). While in this latter case the larger particles tend to accumulate in the warmer region, there is a change in the sign of $\Lambda$ at a given value of the concentration in the other cases analyzed. In particular, the larger species tend to move towards the colder plate when they experience more inelastic collisions than the other ones ($\alpha_{11}<\alpha_{22}$). Moreover, although not shown in the figure, our results also indicate a very weak influence of the volume fraction $\phi$ on the segregation process in this special case.

\section{Phase diagrams for the BNE/RBNE transition}
\label{sec4}

\begin{figure}
\includegraphics[width=0.5 \columnwidth,angle=0]{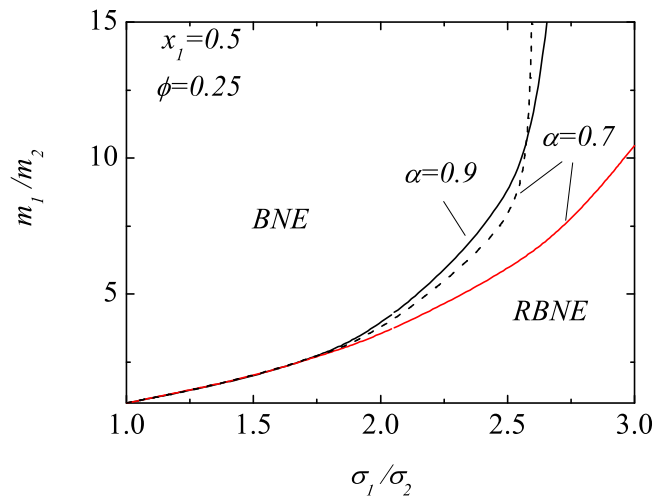}
\caption{(color online) BNE/RBNE phase diagram for inelastic hard spheres ($d=3$) with $x_1=1/2$, $\phi=0.25$, and two values of the (common) coefficient of restitution $\alpha\equiv \alpha_{ij}$. Points above the curve correspond to $\Lambda>0$ (BNE) while points below the curve correspond to $\Lambda<0$ (RBNE). The solid lines are the results derived from Eq.\ \eqref{3.10} while the dashed line is the result obtained from Eq.\ \eqref{3.10} for $\alpha=0.7$ but assuming energy equipartition ($T_1=T_2$).
\label{fig5}}
\end{figure}
\begin{figure}
\includegraphics[width=0.5 \columnwidth,angle=0]{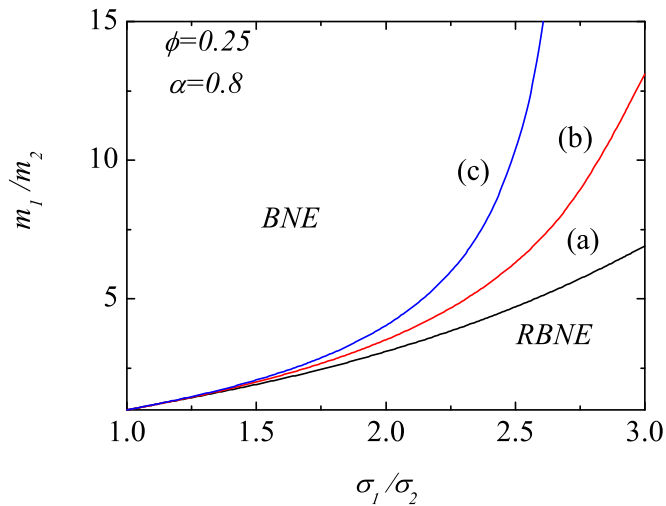}
\caption{(color online) BNE/RBNE phase diagram for inelastic hard spheres ($d=3$) with $\phi=0.25$, $\alpha\equiv \alpha_{ij}=0.8$, and three different values of the concentration $x_1$: (a) $x_1=0.1$, (b) $x_1=0.4$, and (c) $x_1=0.7$ . Points above the curve correspond to $\Lambda>0$ (BNE) while points below the curve correspond to $\Lambda<0$ (RBNE).
\label{fig6}}
\end{figure}

Beyond the special limit situations considered in the previous Section, the thermal diffusion factor $\Lambda$ [or, equivalently, the segregation criterion (\ref{2.17})] depends in
general on the following dimensionless parameters: the mass ratio $m_1/m_2$, the diameter ratio $\sigma_1/\sigma_2$, the concentration $x_1$, the overall volume fraction $\phi$, and the coefficients of restitution $\alpha_{11}$, $\alpha_{22}$ and $\alpha_{12}$. For purposes of simplicity, henceforth the coefficients of restitution will be assumed to be the same for all combinations of collisions (i.e., $\alpha_{11}=\alpha_{22}=\alpha_{12}\equiv \alpha$). Moreover, we only consider the physical case of hard spheres ($d=3$). This reduces the parameter space to five parameters.

\begin{figure}
\includegraphics[width=0.5 \columnwidth,angle=0]{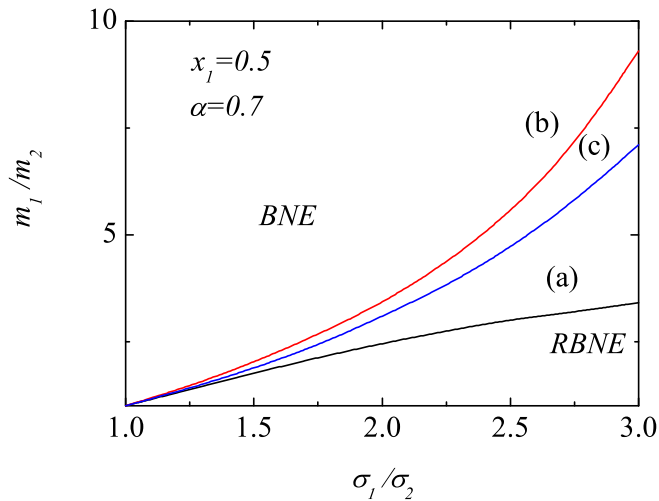}
\caption{(color online) BNE/RBNE phase diagram for an equimolar mixture ($x_1=0.5$) of inelastic hard spheres ($d=3$)  with $\alpha\equiv \alpha_{ij}=0.7$ and three different values of the solid volume fraction $\phi$: (a) $\phi=0.1$, (b) $\phi=0.2$, and (c) $\phi=0.4$ . Points above the curve correspond to $\Lambda>0$ (BNE) while points below the curve correspond to $\Lambda<0$ (RBNE).
\label{fig7}}
\end{figure}
Next, we illustrate the form of the phase diagrams delineating the regimes between BNE and RBNE in the $\{\sigma_1/\sigma_2, m_1/m_2\}$-plane as functions of the concentration $x_1$, the overall volume fraction $\phi$ and the (common) coefficient of restitution $\alpha$. First, Fig.\ \ref{fig5} shows a phase diagram for an equimolar mixture ($x_1=0.5$) at $\phi=0.25$ (moderate density). Two different values of $\alpha$ have been considered ($\alpha=0.9$ and 0.7). In contrast to what happens in the intruder limit case (see, for instance, Fig.\ 4 of Ref.\ \cite{G09}), it is apparent that the main effect of collisional dissipation is to reduce the size of the RBNE region. We observe that in general the RBNE (large particles tend to move towards the hot regions) is dominant for both small mass ratio and/or large size ratio. Moreover, in order to gauge the impact of the non-equipartition of granular energy on segregation, we have also included in Fig.\ \ref{fig5} the phase diagram for $\alpha=0.7$ obtained from the segregation criterion \eqref{2.17} but assuming energy equipartition ($T_1=T_2$). This has been a usual simplification in many previous theoretical works on thermal diffusion in nearly elastic systems \cite{AW98,JY02}. The comparison indicates a good qualitative agreement (at least in the region shown in the figure) between both results for not too large size ratios. On the other hand, quantitative discrepancies appear as the size ratio increases. In particular, although not shown in the phase diagram, when $T_1=T_2$ the mass ratio becomes a two-valued function of the size ratio for values of $\sigma_1/\sigma_2\gtrsim 2.6$. This means that there exists a threshold value of $\sigma_1/\sigma_2$ above which no BNE is predicted. It must be remarked that the significant influence of the different partial temperatures $T_i$ on  thermal diffusion found here is consistent with the molecular dynamics findings of Galvin {\em et al.} \cite{GDH05}. These authors showed that non-equipartition driving forces for segregation are comparable to other driving forces for systems displaying comparable level of non-equipartition. Regarding this point it must be remarked that for systems where segregation is mainly driven by gravity (molecular dynamics simulations of Ref.\ \cite{GDH05} were carried out in the absence of gravity), previous theoretical results \cite{JY02,YJ06,G09} have clearly shown that non-equipartition has a weaker influence on segregation for thermalized systems (i.e., when $\partial_zT\to 0$) than in the opposite limit (absence of gravity). This behavior qualitatively agrees with the experiments carried out by Schr\"oter {\em et al.} \cite{SUKSS06}.

Let us now analyze the effect of the concentration $x_1$ of the large particles on segregation. This is one of the main added values of the present paper with respect to previous studies focused on the tracer limit case ($x_1\to 0$). Figure \ref{fig6} shows a phase diagram for $\alpha=0.8$, $\phi=0.25$, and three different values of the mole fraction $x_1$. We observe that the concentration of the mixture has significant effects in reducing the BNE region as $x_1$ increases. In particular, for a given value of the concentration, the transition from BNE to RBNE may occur following two paths: i) along constant mass ratio $m_1/m_2$ with increasing size ratio $\sigma_1/\sigma_2$, and ii) along constant size ratio $\sigma_1/\sigma_2$ with decreasing mass ratio $m_1/m_2$. Finally, Fig.\ \ref{fig7} illustrates the influence of the volume fraction on the phase diagram for an equimolar mixture ($x_1=0.5$) at a moderate level of collisional dissipation ($\alpha=0.7$). It is apparent that the role played by the density is quite important, especially for large size ratios. Surprisingly, and in contrast to the intruder limit case (see for instance Fig.\ 5 of Ref.\ \cite{G09}), the dependence on the solid fraction $\phi$ is not monotonic: while the range of size and mass ratios for which the RBNE exists decreases with decreasing $\phi$ at moderate densities (lines corresponding to $\phi=0.1$ and 0.2), the opposite happens at higher densities. Thus, at a given value of the size ratio, one has  $M^{\text{cr}}(0.1)<M^{\text{cr}}(0.4)<M^{\text{cr}}(0.2)$, where $M^{\text{cr}}(\phi)$ denotes the critical mass ratio where the transition from RBNE to BNE occurs at density $\phi$.

\section{Comparison with molecular dynamics simulations}
\label{sec5}

To the best of our knowledge, one of the few molecular dynamics simulations in which thermal diffusion has been isolated from the remaining segregation mechanisms has been reported by Galvin {\em et al.} \cite{GDH05}. They consider a binary granular mixture constituted by frictionless inelastic spheres subject to an imposed temperature gradient. As expected, their results show in general segregation of particles according to their size and/or density (species segregation). Since these authors are mainly interested in assessing the role played by the non-equipartition of energy on segregation, no external forces like gravity are present in their simulations and hence the system is characterized by zero mean flow.

In the steady state, the granular temperature gradient between walls drives the segregation process. As in other experiments \cite{BRM05,WJKT06}, the temperature profile demonstrates nonlinear behavior and exhibits a global minimum near the cold wall. In addition, Galvin {\em et al.} \cite{GDH05} also examine the profiles of solid volume fraction $\phi(z)$ and partial densities $n_i(z)$ across the system to provide a quantitative measure of segregation (see, for instance, Fig.\ 7 of Ref.\ \cite{GDH05}).

Although a direct comparison between the theoretical results derived here with those obtained in Ref.\ \cite{GDH05} for the partial density profiles $n_i(z)$ would require the numerical solution of the condition $j_{1,z}=0$ along with the energy balance equation \eqref{2.3} (in its steady state version), we will restrict our comparison to the thermal diffusion factor $\Lambda$. In fact, this quantity provides a more qualitative property of $n_i$ since its sign gives the tendency of each species to move upwards or downwards. In order to make a close comparison between theory and simulation for the thermal diffusion, let us consider the simulation data reported in Fig.\ 7 of Ref.\ \cite{GDH05} for a binary mixture with mechanical parameters $\sigma_1/\sigma_2=2$, $m_1/m_2=16$ and a (common) coefficient of restitution $\alpha=0.9$. According to the results displayed in this figure, it is quite apparent that $\Lambda$ is a {\em nonuniform} function since it depends on $z$ through its dependence on the volume fraction $\phi(z)$ and the concentration $x_1(z)=n_1(z)/n(z)$. Therefore, one can determine $\Lambda(\phi(z),x_1(z))$ from Eq.\ \eqref{2.16} by using the local  values of $\phi(z)$ and $x_1(z)$ provided by the simulation as input parameters. This will give us the theoretical prediction of thermal diffusion across the system. The corresponding (local) value for $\Lambda$ predicted by the simulations can be estimated from the temperature and species density profiles by numerically computing the derivatives $\partial_z\ln (n_1/n_2)$ and $\partial_z \ln T$ at each point of the system. In this case, according to Eq.\ \eqref{2.8}, the value $\Lambda_{\text{MD}}$ given by the simulation is
\begin{equation}
\label{5.1}
\Lambda_{\text{MD}}=-\frac{\partial_z\ln
\left(n_1/n_2\right)|_\text{MD}}{\partial_z \ln T|_\text{MD}},
\end{equation}
where the subscript $\text{MD}$ means that these derivatives are obtained from the simulation data.

The theoretical and molecular dynamics simulation results for $\Lambda$ are shown in Table \ref{tab1}. Regarding the sign of $\Lambda$, the comparison between theory and simulation shows a good agreement since both predict positive values for $\Lambda$ in all the range of parameters $(\phi, x_1)$ analyzed. Consequently, the more massive particles segregate preferentially toward the cool region, in qualitative agreement with the snapshot shown in Fig.\ 4 of Ref.\ \cite{GDH05}.
At a more quantitative level, although theory and simulation compare well in the case of points near the minimum of temperature, there are in general discrepancies between theory and simulation. This quantitative disagreement can be due to the fact that while the expression \eqref{2.16} for $\Lambda$ has been obtained up to the Navier-Stokes order (first order in the spatial gradients), the molecular dynamics simulations carried out in Ref.\ \cite{GDH05} clearly show strong gradients in solid fraction.
Presumably, the numerical solution (beyond the Navier-Stokes domain) of the Enskog equation via the Direct Simulation Monte Carlo method \cite{B94} would give a better quantitative agreement with molecular dynamics simulations than the Navier-Stokes results reported here. This is a quite interesting problem to be addressed in the near future.

\begin{table}[tbp]
\begin{tabular}{|c|c|c|c|}
\cline{1-4}
$\phi$ & $x_1$ & $\Lambda_{\text{MD}}$ & $\Lambda$\\
\cline{1-4}
0.33 & 0.43 & 1.09 & 0.60 \\
0.45 & 0.63 & 1.21 & 1.25 \\
0.50 & 0.70 & 1.02 & 1.71 \\
0.51 & 0.71 & 0.83 & 1.76 \\
0.48 & 0.68 & 0.90 & 1.49 \\
0.41 & 0.58 & 1.03 & 0.96 \\
0.30 & 0.39 & 1.26 & 0.54 \\
0.20 & 0.21& 1.80 & 0.47 \\
0.10 & 0.07 & 2.44 & 1.22 \\
0.08 & 0.04 & 2.29 & 1.79 \\
0.06 & 0.03 &  2.19 & 2.39 \\
\cline{1-4}
\end{tabular}
\caption{Thermal diffusion factor as obtained from molecular dynamics simulations ($\Lambda_{\text{MD}}$) and the Enskog theory ($\Lambda$) for different values of volume fraction $\phi$ and concentration $x_1$ for a granular binary mixture constituted by spheres with $\sigma_1/\sigma_2=2$, $m_1/m_2=16$, and $\alpha=0.9$. Molecular dynamics results have been obtained by Galvin {\em et al.} \cite{GDH05}.
\label{tab1}}
\end{table}

\section{Summary and discussion}
\label{sec6}

The understanding of particle segregation within polydisperse, rapid granular flows is still not well understood. The reason is perhaps twofold: first, there is a large number of relevant parameters involved in the description of the granular mixture; and second, there is a wide array of complexities that arise during the derivation of kinetic theory models. As previously mentioned, the two most common simplifications used in previous theoretical works on segregation have been to consider systems constituted by nearly elastic particles and an equipartition of granular energy. This paper has addressed the problem of segregation by thermal diffusion in a binary granular mixture described by the inelastic Enskog equation. The analysis has been based on a solution of the Enskog equation that covers some of the aspects not accounted for in previous studies. Specifically, i) it takes into account the nonlinear dependence of the transport coefficients on collisional dissipation (and thus the theory is expected to be applicable for a wide range of coefficients of restitution); ii) it considers the influence of the nonequipartition of granular energy on segregation; and iii) it applies to moderate densities. Consequently, the theory subsumes all previous analysis for both dilute \cite{SGNT06,SNTG09,BRM05, G06} and dense \cite{AW98,HH96,JY02,TAH03} gases, which are recovered in the appropriate limits. The results presented here generalize to \emph {arbitrary} concentration previous results \cite{G08,G09,GV09} obtained in the tracer limit ($x_1\to 0$).

Among the different mechanisms involved in segregation, thermal diffusion (segregation induced by a temperature gradient) becomes the most relevant one when the granular system behaves like a granular gas. In the steady state with gradients only along a given direction, the \emph{sign} of the thermal diffusion factor $\Lambda$ [defined by Eq.\ \eqref{2.8}] provides information on the tendency of each species to move towards the colder or hotter plate. In this paper, the factor $\Lambda$ has been evaluated by following two complementary approaches. First, by using the momentum balance equation \eqref{2.12} (in the absence of gravity) along with the constitutive equation \eqref{2.4} for the mass flux, $\Lambda$ has been expressed [see Eq.\ \eqref{2.16}] in terms of the pressure $p$ (and its derivatives with respect to the partial densities $n_i$) and the transport coefficients $D_{11}$ , $D_{12}$, and $D_1^T$ associated with the mass flux. Then, the forms of the pressure and the diffusion transport coefficients have been explicitly obtained from a Chapman-Enskog solution of the Enskog equation \cite{GDH07,GHD07}. This finally gives $\Lambda$ as a function of the mass and size ratios, the concentration, the solid volume fraction, and the coefficients of restitution. In particular, the condition $\Lambda=0$ [see Eq.\ \eqref{2.17}] provides the segregation criterion for the transition BNE$\Leftrightarrow$ RBNE.

In general, the segregation criterion \eqref{2.17} presents a complex dependence on the parameter space of the system. In order to disentangle the impact of the different parameters on thermal diffusion segregation, some special cases (dilute gas, tracer limit, $\ldots$) have been separately studied. An interesting new case corresponds to the segregation of a binary mixture of granular particles that differ \emph{only} by their coefficients of restitution. This novel effect was first predicted by Serero  \emph{et al}. \cite{SGNT06,SNTG09} from the Boltzmann equation (low-density gas) and has been recently confirmed by molecular dynamics simulations of hard disks \cite{BEGS08}. The results obtained here for dense granular binary mixtures confirm also the existence of segregation induced by an inelasticity difference (see Fig.\ \ref{fig4}). Moreover, our results also show in general a weak influence of the volume fraction on thermal diffusion for this special situation.

A systematic study of the form of the phase diagrams BNE/RBNE in the mass and size ratio plane has been carried out in Section \ref{sec4} for hard spheres in the case $\alpha_{ij}=\alpha$. Regarding the influence of collisional dissipation on the form of the phase diagrams, the results indicate that the influence of $\alpha$ on $\Lambda$ is quite important, the main effect of dissipation being to increase the size of the BNE region (see Fig.\ \ref{fig5}). In addition, we also conclude that the role played by the  nonequipartition of granular energy on segregation is quite relevant, especially as the disparity of masses and/or sizes increases. This result is consistent with recent molecular dynamics simulations \cite{GDH05}. With respect to the influence of the concentration $x_1$, our results show that in general the main effect of $x_1$ is to reduce the BNE region as the concentration of the large particles increases (see Fig.\ \ref{fig6}). Finally, we also observe that the form of the phase diagrams changes significantly with the volume fraction $\phi$, specially at large size ratios (see Fig.\ \ref{fig7}).

By extending the intruder limit analysis \cite{G08,G09}  to arbitrary values of concentration $x_1$, comparisons with molecular dynamics simulations become practical and this allows one to assess the reliability of the Enskog kinetic theory to characterize thermal diffusion segregation. To make some contact with the molecular dynamics results of Galvin \emph{et al.} \cite{GDH05}, we have compared the kinetic theory predictions for the thermal diffusion factor $\Lambda$ for different values of the concentration $x_1$ and volume fraction $\phi$ (for the system $\sigma_1/\sigma_2=2$, $m_1/m_2=16$ and $\alpha=0.9$) with those obtained from the simulation data by numerically evaluating the derivatives $\partial_z\ln (n_1/n_2)$ and $\partial_z \ln T$ at the points of the system corresponding to the same values of $x_1$ and $\phi$. The comparison between theory and simulation shows a good qualitative agreement since both predict the same sign of $\Lambda$ for the different points analyzed. In addition, at a more quantitative level, although molecular dynamics simulations show {\em strong} gradients in the bulk region (and so they go beyond the linear domain of the Navier-Stokes description), the theory compares reasonably well with simulation, especially in the region close to the minimum of granular temperature. It is important to remark again that the quantitative discrepancies between theory and molecular dynamics  simulations are essentially due to the limitations of the Navier-Stokes results rather than the assumptions inherent to the Enskog kinetic equation (molecular chaos hypothesis).

Certainly, the derivation of kinetic theory models for segregation flows in polydisperse systems is perhaps one of the most important open challenges of granular gas research. The theoretical results reported in this paper cover part of this challenge, at least in the case of the thermal diffusion segregation. On the other hand, the present theory has some important restrictions. First, given that the Enskog equation still assumes uncorrelated particle velocities (molecular chaos hypothesis), it is expected that the kinetic theory for thermal diffusion only applies to \emph{moderate} densities (solid volume fraction typically smaller than or equal to 0.25). However, despite this limitation, there is substantial evidence in the literature \cite{Enskog} on the reliability of the Enskog theory to accurately describe macroscopic properties (such as transport coefficients) for a wide range of densities and/or collisional dissipation. Another important limitation is that the segregation criterion derived here has been obtained by using the first Sonine approximation for the diffusion transport coefficients. Recent results \cite{GV09} for the tracer limit clearly show that the accuracy of the first Sonine solution can be questionable for small values of the coefficients of restitution and/or disparate values of the mass and size ratios. The influence of the second Sonine correction to the transport coefficients is an interesting open problem to be carried out in the near future. This will allow us to offer a segregation theory that can be reliable even for extreme values of dissipation or mass and size ratios.

\acknowledgments

I am quite grateful to Janine Galvin for evaluating the thermal diffusion factor from the simulation data of Fig.\ 7 of Ref.\ \cite{GDH05}. The present work has been supported by the Ministerio de Ciencia e Innovaci\'on  (Spain) through grant No. FIS2010-16587, partially financed by
FEDER funds and by the Junta de Extremadura (Spain) through Grant No. GRU10158.

\appendix
\section{Expressions of the mass flux transport coefficients and pressure}
\label{appA}

In this Appendix we provide the expressions of the (reduced) transport coefficients $D_{11}^*$, $D_{12}^*$, and $D_{1}^{T*}$ associated with the mass flux and the hydrostatic pressure $p^*$. These quantities are involved in the evaluation of the thermal diffusion factor $\Lambda$.

The expressions of the reduced coefficients $D_{1}^{T*}$, $D_{11}^*$,
and $D_{12}^*$ can be written as \cite{AHG10}
\begin{eqnarray}
D_{1}^{T*}&=&\left(\nu_D^*-\zeta^*\right)^{-1}\Big\{x_1\gamma_1-\frac{p^*\rho_1}{\rho}+\frac{\pi ^{d/2}}{2d\Gamma \left( \frac{d}{2}\right)}x_1n\sigma_2^{d}\left[x_1\chi_{11}(\sigma_1/\sigma_2)^{d}\gamma_1(1+\alpha_{11})\right.
\nonumber\\
& &
\left.
+2x_2\chi_{12}(\sigma_{12}/\sigma_2)^{d}M_{12}\gamma_2(1+\alpha_{12})\right]\Big\},  \label{a1}
\end{eqnarray}
\begin{eqnarray}
\label{a2}
\left(\nu_D^*-\frac{1}{2}\zeta^*\right)D_{11}^*&=&\frac{D_{1}^{T*}}{x_1\nu_0}n_1\frac{\partial \zeta^{(0)}}{\partial n_1}-\frac{m_1}{\rho T}n_1\frac{\partial p}{\partial n_1}+\gamma_1+n_1\frac{\partial \gamma_1}{\partial n_1}\nonumber\\
& & +\frac{\pi^{d/2}}{d\Gamma \left( \frac{d}{2}\right)}
x_1 n\sigma_2^{d}\sum_{\ell=1}^{2}\chi_{1\ell}(\sigma _{1\ell}/\sigma_2)^{d}M_{\ell 1}(1+\alpha_{1\ell })
\nonumber\\
& & \times
\left\{ \frac{1}{2}\left(\gamma_1+\frac{m_1}{m_{\ell}}\gamma_\ell\right)\left[ 2\delta _{1\ell }+n_\ell
\frac{\partial \ln \chi_{1\ell}}{\partial n_1}+\frac{n_\ell}{n_1}I_{1\ell 1}\right]+\frac{m_1}{m_\ell}n_\ell
\frac{\partial \gamma_\ell}{\partial n_1}\right\},
\nonumber\\
\end{eqnarray}
\begin{eqnarray}
\label{a3}
\left(\nu_D^*-\frac{1}{2}\zeta^*\right)D_{12}^*&=&\frac{D_{1}^{T*}}{(1-x_1)\nu_0}n_2\frac{\partial \zeta^{(0)}}{\partial n_2}-\frac{m_1}{\rho T}n_1\frac{\partial p}{\partial n_2}+n_1
\frac{\partial \gamma_1}{\partial n_2}\nonumber\\
& & +\frac{\pi^{d/2}}{d\Gamma \left( \frac{d}{2}\right)}
x_1 n\sigma_2^{d}\sum_{\ell=1}^{2}\chi_{1\ell}(\sigma _{1\ell}/\sigma_2)^{d}M_{\ell 1}(1+\alpha_{1\ell })
\nonumber\\
& & \times
\left\{ \frac{1}{2}\left(\gamma_1+\frac{m_1}{m_{\ell}}\gamma_\ell\right)\left[ 2\delta _{2\ell }+n_\ell
\frac{\partial \ln \chi_{1\ell}}{\partial n_2}+\frac{n_\ell}{n_2}I_{1\ell 2}\right]+\frac{m_1}{m_\ell}n_\ell
\frac{\partial \gamma_\ell}{\partial n_2}\right\}.
\nonumber\\
\end{eqnarray}
In these equations, $\gamma_i=T_i/T$, $\zeta^*=\zeta^{(0)}/\nu_0$, $p^*=p/(nT)$, $\chi_{ij}$ is the pair distribution function at contact, $M_{ij}=m_i/(m_i+m_j)$, and
\begin{equation}
\label{a4}
\nu_D^*= \frac{2\pi^{(d-1)/2}}{d\Gamma \left( \frac{d}{2}\right)}\chi_{12}(1+\alpha_{12})\left(\frac{\theta_1+\theta_2}
{\theta_1\theta_2}\right)^{1/2}(x_1 M_{12}+x_2 M_{21}),
\end{equation}
where $\theta_i=m_iT/m_0T_i$ and $m_0\equiv (m_1+m_2)/2$.
The partial temperatures $T_1$ and $T_2$ are determined from the condition
$\zeta_1^{(0)}=\zeta_2^{(0)}=\zeta^{(0)}$, where the expression of $\zeta_i^{(0)}$ is
\begin{eqnarray}
\label{a5}
\zeta^{(0)}=\zeta_i^{(0)}&=&\frac{4\pi^{(d-1)/2}}{d\Gamma \left(\frac{d}{2}\right)}\nu_0\sum_{j=1}^2\chi_{ij}x_jM_{ji}(\sigma_{ij}/\sigma_{12})^{d-1}
\left(\frac{\theta_i+\theta_j}{\theta_i\theta_j}\right)^{1/2}
\left(1+\alpha_{ij}\right)\nonumber\\
& & \times \left[1-\frac{M_{ji}}{2}\left(1+\alpha_{ij}
\right)\frac{\theta_i+\theta_j}{\theta_j}\right],
\end{eqnarray}
where $\nu_0=n\sigma_{12}^{d-1}\sqrt{2T/m_0}$. The reduced pressure $p^*$ is given by \cite{GDH07}
\begin{equation}
\label{a10}
p^*=1+\frac{\pi^{d/2}}{d\Gamma \left(\frac{d}{2}\right)}n\sigma_2^d\sum_{i=1}^2\sum_{j=1}^2\;x_ix_j(\sigma_{ij}/\sigma_2)^d
M_{ji}\left(1+\alpha_{ij}\right)\chi_{ij}\gamma_i.
\end{equation}

The explicit form of the transport coefficients $D_{11}^*$ and $D_{12}^*$ requires the knowledge of the the quantities $I_{i\ell j}$. These parameters are given in terms of the functional derivative of the (local) pair distribution function $\chi_{ij}$ with respect to the (local) partial densities $n_\ell$
[see Eq.\ (C11) of Ref.\ \cite{GDH07}]. Given the mathematical difficulties involved in the determination of the above functional derivatives, for the sake of simplicity, the parameters $I_{i\ell j}$ are chosen here to recover the results derived by L\'opez de Haro {\em et al.} for elastic mixtures \cite{MCK83} (see Appendix C of Ref.\ \cite{GHD07}). The
quantities $I_{i\ell j}$ are the origin of the primary difference between the
standard Enskog theory and the revised version for elastic collisions \cite{BE73}. They are zero if
$i=\ell$, but otherwise are not zero. These quantities are defined through the relation \cite{GHD07}
\begin{equation}
\label{a7} \sum_{\ell=1}^2 n_\ell\sigma_{i\ell}^d \chi_{i\ell}\left(n_j
\frac{\partial\ln \chi_{i\ell}}{\partial n_j}+I_{i\ell j}\right)=
\frac{n_j}{TB_2}\left(\frac{\partial \mu_i}{\partial n_j}\right)_{T,n_{k\neq
j}}-\frac{\delta_{ij}}{B_2}-2n_j\chi_{ij}\sigma_{ij}^d,
\end{equation}
where $\mu_i$ is the chemical potential of species $i$ and $B_2=\pi^{d/2}/d\Gamma(d/2)$ [$B_2=\frac{\pi}{2}$ for disks ($d=2$) and $B_2=\frac{2\pi}{3}$ for spheres ($d=3$)].
Since granular fluids lack a thermodynamic description, the concept of chemical potential could be questionable. As said before, the presence of $\mu_i$ in our theory is essentially due to the choice of the quantities $I_{i\ell j}$. Given that the explicit form of the chemical potential must be known to evaluate the diffusion transport coefficients, for practical purposes, the expression considered here for $\mu_i$ is the same as the one obtained for an ordinary mixture of gases ($\alpha_{ij}=1$). Although this evaluation requires the use of thermodynamic relations that only apply for elastic systems, we expect that this approximation could be reliable for not too strong values of dissipation. More comparisons with computer simulations are needed to support the above expectation.

Taking into account Eq.\ (\ref{a7}), the nonzero parameters $I_{121}$ and $I_{122}$ appearing in Eqs.\ \eqref{a2} and \eqref{a3} are given by
\begin{equation}
\label{a8}
I_{121}=\frac{1}{TB_2n_2\sigma_{12}^d\chi_{12}}\left[n_1\left(\frac{\partial \mu_1}{\partial n_1}
\right)_{T,n_{2}}-T\right]-2\frac{n_1\sigma_1^d\chi_{11}}
{n_2\sigma_{12}^d\chi_{12}}-
\frac{n_1^2\sigma_1^d}{n_2\sigma_{12}^d\chi_{12}}\frac{\partial\chi_{11}}{\partial n_1}
-\frac{n_1}{\chi_{12}}\frac{\partial\chi_{12}}{\partial n_1},
\end{equation}
\begin{equation}
\label{a9}
I_{122}=\frac{1}{TB_2\sigma_{12}^d\chi_{12}}\left(\frac{\partial \mu_1}{\partial n_2}
\right)_{T,n_{1}}-2-
\frac{\sigma_1^d n_1}{\sigma_{12}^d\chi_{12}}\frac{\partial\chi_{11}}{\partial n_2}
-\frac{n_2}{\chi_{12}}\frac{\partial\chi_{12}}{\partial n_2}.
\end{equation}
Note that for mechanically equivalent particles ($m_1=m_2$, $\sigma_1=\sigma_2$), $I_{121}=I_{122}=0$, as expected since the standard and revised versions of the Enskog equation lead to the same Navier-Stokes transport coefficients for a monocomponent gas \cite{BE73,MG93}.

In the case of hard disks ($d=2$), a good approximation for the pair distribution function $\chi_{ij}$ is \cite{JM87}
\begin{equation}
\label{a15}
\chi_{ij}=\frac{1}{1-\phi}+\frac{9}{16}\frac{\phi}{(1-\phi)^2}\frac{\sigma_i\sigma_jM_1}{\sigma_{ij}M_2},
\end{equation}
where $\phi=\sum_i\; n_i\pi \sigma_i^2/4$ is the solid volume fraction for disks and
\begin{equation}
\label{a16}
M_n=\sum_{s=1}^2\; x_s \sigma_s^n.
\end{equation}
The expression of the chemical potential $\mu_i$ of the species $i$
consistent with the approximation (\ref{a15}) is \cite{S08}
\begin{eqnarray}
\label{a17}
\frac{\mu_i}{T}&=&\ln (\lambda_i^2n_i)-\ln (1-\phi)+\frac{M_1}{4M_2}\left[\frac{9\phi}{1-\phi}+\ln (1-\phi)\right]\sigma_i
\nonumber\\
& &
-\frac{1}{8}\left[\frac{M_1^2}{M_2^2}\frac{\phi(1-10\phi)}{(1-\phi)^2}-
\frac{8}{M_2}\frac{\phi}{1-\phi}+\frac{M_1^2}{M_2^2}\ln (1-\phi)\right]\sigma_i^2,
\end{eqnarray}
where $\lambda_i(T)$ is the (constant) de Broglie's thermal wavelength \cite{RG73}. In the case of hard spheres ($d=3$), we take for the pair distribution function
$\chi_{ij}$ the following approximation \cite{B70}
\begin{equation}
\label{a18}
\chi_{ij}=\frac{1}{1-\phi}+\frac{3}{2}\frac{\phi}{(1-\phi)^2}\frac{\sigma_i\sigma_jM_2}{\sigma_{ij}M_3}
+\frac{1}{2}\frac{\phi^2}{(1-\phi)^3}\left(\frac{\sigma_i\sigma_jM_2}{\sigma_{ij}M_3}\right)^2,
\end{equation}
where $\phi=\sum_i\; n_i\pi \sigma_i^3/6$ is the solid volume fraction for spheres.
The chemical potential consistent with (\ref{a18}) is \cite{RG73}
\begin{eqnarray}
\label{a19}
\frac{\mu_i}{T}&=&\ln (\lambda_i^3n_i)-\ln (1-\phi)+3\frac{M_2}{M_3}\frac{\phi}{1-\phi}\sigma_i
+3\left[\frac{M_2^2}{M_3^2}\frac{\phi}{(1-\phi)^2}+
\frac{M_1}{M_3}\frac{\phi}{1-\phi}+\frac{M_2^2}{M_3^2}\ln (1-\phi)\right]\sigma_i^2\nonumber\\
& & -\left[\frac{M_2^3}{M_3^3}\frac{\phi(2-5\phi+\phi^2)}{(1-\phi)^3}-3
\frac{M_1M_2}{M_3^2}\frac{\phi^2}{(1-\phi)^2}-\frac{1}{M_3}\frac{\phi}{1-\phi}
+2\frac{M_2^3}{M_3^3}\ln (1-\phi)\right]\sigma_i^3.
\end{eqnarray}

According to Eqs.\ \eqref{a1}--\eqref{a3}, the diffusion transport coefficients are given in terms of the derivatives of $\gamma_i$ with respect to the partial densities $n_i$. In terms of the temperature ratio $\gamma=T_1/T_2$, the partial temperatures $\gamma_i$ are defined as
\begin{equation}
\label{a20}
\gamma_1=\frac{\gamma}{1+x_1(\gamma-1)}, \quad
\gamma_2=\frac{1}{1+x_1(\gamma-1)}.
\end{equation}
The dependence of the temperature ratio $\gamma$ on $n_1$ and $n_2$ is through its dependence on the concentration $x_1$ and the volume fraction $\phi$. As a consequence,
\begin{equation}
\label{a21}
n_1\frac{\partial \gamma}{\partial n_1}=\phi_1 \frac{\partial \gamma}{\partial \phi}
+x_1(1-x_1)\frac{\partial \gamma}{\partial x_1},
\end{equation}
\begin{equation}
\label{a22}
n_2\frac{\partial \gamma}{\partial n_2}=\phi_2 \frac{\partial \gamma}{\partial \phi}
-x_1(1-x_1)\frac{\partial \gamma}{\partial x_1},
\end{equation}
where $\phi_i$ is defined by Eq.\ \eqref{2.14.3}. The derivatives $\partial_\phi\gamma$ and $\partial_{x_1} \gamma$ can be obtained by taking the derivatives with respect to $\phi$ and $x_1$ in the condition $\zeta_1^*=\zeta_2^*$. This yields the relations
\begin{equation}
\label{a23}
\frac{\partial \gamma}{\partial \phi}=
\frac{\left(\frac{\partial \zeta_1^*}{\partial \phi}\right)_{\gamma}-\left(\frac{\partial \zeta_2^*}{\partial \phi}\right)_{\gamma}}{\left(\frac{\partial \zeta_2^*}{\partial \gamma}\right)-\left(\frac{\partial \zeta_1^*}{\partial \gamma}\right)}, \quad \frac{\partial \gamma}{\partial x_1}=
\frac{\left(\frac{\partial \zeta_1^*}{\partial x_1}\right)_{\gamma}-\left(\frac{\partial \zeta_2^*}{\partial x_1}\right)_{\gamma}}{\left(\frac{\partial \zeta_2^*}{\partial \gamma}\right)-\left(\frac{\partial \zeta_1^*}{\partial \gamma}\right)}.
\end{equation}
This allows us to express the derivatives $\partial_\phi\gamma$ and $\partial_{x_1} \gamma$ in terms of the temperature ratio $\gamma$ and the parameters of the mixture.

It is apparent that the (reduced) transport coefficients have a complex dependence on the mass and size ratios, the concentration, the volume fraction, and the coefficients of restitution. A simple but nontrivial case corresponds to a binary system constituted by mechanically equivalent particles ($m_1=m_2$, $\sigma_1=\sigma_2$, $\alpha_{11}=\alpha_{22}=\alpha_{12}$). In this case, $\chi_{ij}=\chi$, $\gamma_i=\gamma=1$, $p^*=1+2^{d-2}\chi\phi (1+\alpha)$, and the parameters $I_{ij\ell}=0$. As a consequence, Eqs.\ \eqref{a1}--\eqref{a3} simply reduce to $D_1^{T*}=0$ and
\begin{equation}
\label{a24}
D_{12}^*=-\frac{x_1}{x_2}D_{11}^*, \quad
D_{11}^*=\left(\nu_D^*-\frac{1}{2}\zeta^*\right)^{-1}\left[1-x_1(p^*+\phi\partial_\phi p^*)
+2 x_1(p^*-1)\left(1+\phi \partial_\phi \ln \chi \right)\right],
\end{equation}
where
\begin{equation}
\label{a24.1}
\nu_D^*=\frac{\sqrt{2}\pi^{(d-1)/2}}{d\Gamma \left(\frac{d}{2}\right)}\chi (1+\alpha), \quad
\zeta^*=\frac{\sqrt{2}\pi^{(d-1)/2}}{d\Gamma \left(\frac{d}{2}\right)}\chi (1-\alpha^2).
\end{equation}

\section{Tracer limit case}

The explicit forms of the diffusion transport coefficients in the tracer limit case ($x_1\to 0$) are displayed in this Appendix. In this limit, $\gamma_2=1$, $\gamma_1=\gamma(\phi)$ and $p^*=1+2^{d-2}\chi_{22}\phi (1+\alpha_{22})$. Moreover, since the dependence of $ \zeta^{(0)}$, $p$ and $\gamma$ on the partial densities is only through the volume fraction $\phi$, one gets the simple relations $\partial_{n_1}\zeta^{(0)}=\partial_{n_1}p=\partial_{n_1}\gamma=0$,
\begin{equation}
\label{b1}
n_2\frac{\partial \zeta^{(0)}}{\partial n_2}=\zeta^{(0)}\left(1+\phi\frac{\partial \ln \chi_{22}}{\partial \phi}\right),
\end{equation}
\begin{equation}
\label{b2}
\frac{\partial p}{\partial n_2}=p^*\left(1+\phi\frac{\partial \ln p^*}{\partial \phi}\right),
\end{equation}
\begin{equation}
\label{b3}
n_1\frac{\partial \gamma}{\partial n_2}=x_1\phi\frac{\partial \gamma}{\partial \phi}.
\end{equation}
The explicit expressions for the transport coefficients in the tracer limit can be easily obtained from Eqs.\ \eqref{a1}--\eqref{a3} when one takes into account the identities
\eqref{b1}--\eqref{b3}. The result is
\begin{equation}
\label{b4} D_{11}^*=\frac{\gamma}{\nu_D^*-\frac{1}{2}\zeta^*},
\end{equation}
\begin{equation}
\label{b5} D_1^{T*}=x_1\left(\nu_D^*-\zeta^*\right)^{-1}\left[\gamma-M p^*+
\frac{(1+\omega)^d}{2}\frac{M}{1+M} \chi_{12}\phi(1+\alpha_{12})\right],
\end{equation}
\begin{eqnarray}
\label{b6} D_{12}^{*}&=&x_1\left(\nu_D^*-\frac{1}{2}\zeta^*\right)^{-1}\left[\left(1+\phi \partial_\phi \ln \chi_{22} \right)\zeta^*D_1^{T*}-M\beta
+\phi \frac{\partial \gamma}{\partial \phi}\right.\nonumber\\
& & \left.+\frac{1}{2}
\frac{\gamma+M}{1+M}\frac{\phi}{T}\left(\frac{\partial\mu_1}{\partial
\phi}\right)_{T,n_2}(1+\alpha_{12})\right].
\end{eqnarray}
Here, $\beta=p^*+\phi \partial_\phi p^*$, $\mu_1$ is the chemical potential of the tracer particles and
\begin{equation}
\label{b7}
\zeta^*=\frac{\pi^{(d-1)/2}}{d\Gamma \left(\frac{d}{2}\right)}\left(\frac{2}{1+\omega}\right)^{d-1}M_{21}^{-1/2}\chi_{22}
(1-\alpha_{22}^2),
\end{equation}
\begin{equation}
\label{b7}
\nu_D^*=\frac{\sqrt{2}\pi^{(d-1)/2}}{d\Gamma \left(\frac{d}{2}\right)}\chi_{12}
M_{21}^{1/2}\sqrt{\frac{M+\gamma}{M}}(1+\alpha_{12}).
\end{equation}

As said in Sec.\ \ref{sec3}, in order to maintain the granular medium in a fluidized state, previous works \cite{G08,G09} considered the presence of a stochastic external thermostat. The corresponding expressions for the transport coefficients are \cite{note}
\begin{equation}
\label{b8} D_{11}^*=\frac{\gamma}{\nu_D^*},
\end{equation}
\begin{equation}
\label{b9} D_1^{T*}=x_1\nu_D^{*-1}\left[\gamma-M p^*+
\frac{(1+\omega)^d}{2}\frac{M}{1+M} \chi_{12}\phi(1+\alpha_{12})\right],
\end{equation}
\begin{equation}
\label{b10} D_{12}^{*}=x_1\nu_D^{*-1}\left[\phi \frac{\partial \gamma}{\partial \phi}-M\beta+\frac{1}{2}
\frac{\gamma+M}{1+M}\frac{\phi}{T}\left(\frac{\partial\mu_1}{\partial
\phi}\right)_{T,n_1}(1+\alpha_{12})\right].
\end{equation}
Taking into account Eqs.\ \eqref{b8}--\eqref{b10}, the segregation criterion \eqref{3.9}  becomes
\begin{equation}
\label{b11}
\phi\left(p^*\frac{\partial \gamma}{\partial \phi}
-\gamma\frac{\partial p^*}{\partial \phi}\right)+\frac{(1+\omega)^d}{2}M_{12}
\phi\chi_{12}(1+\alpha_{12})\left[\frac{p^*(1+\omega)^{-d}}{T\chi_{12}}\frac{\gamma+M}
{M}\left(\frac{\partial\mu_1}{\partial
\phi}\right)_{T,n_1}-\beta\right]=0.
\end{equation}


\begin{thebibliography} {99}
\bibitem{note}The term proportional to $\partial_\phi \gamma$ in Eq.\ \eqref{b10} was neglected in the expression of $D_{12}^*$ given in Refs.\ \cite{G08} and \cite{G09}. The expression displayed here for $D_{12}^*$ corrects this result. Given that the influence of this term is in general small,
    the general conclusions made in the above references are not altered by the presence of this new term.
\end{thebibliography}
\end{document}